\documentclass{aa} 
\usepackage{graphicx,amssymb,psfig,natbib}

\newcommand{\AaA}{A\&A}

\newcommand{\ApJS}{Astrophysical Journal Supplement Series}
\newcommand{\AJ}{AJ}
\newcommand{\ApJ}{ApJ}

\newcommand{\RvMP}{Reviews of Modern Physics}

\begin{document}

\title{Density structure of the Horsehead nebula photo-dissociation region}

\author{E. Habart\inst{1}, A. Abergel\inst{2},  C.M. Walmsley\inst{1}, D. Teyssier\inst{3} and J. Pety\inst{4,5}}

\institute{Osservatorio Astrofisico di Arcetri, INAF, Largo E. Fermi 5, I-5
0125 Firenze, Italy
\and
Institut d'Astrophysique Spatiale, Universit\'e Paris-Sud, 91405 Orsay, France
\and
Space Research Organization Netherlands, P. O. Box 800, 9700 AV Groningen, The Netherlands (Under an ESA external fellowship) 
\and
LERMA, UMR 8112, CNRS, Observatoire de Paris and Ecole Normale Sup\'erieure, 24 Rue Lhomond, 75231 Paros Cedex 05, France 
\and
IRAM, 300 rue de la Piscine, 38406 Grenoble, cedex, France.}

\abstract{We present high angular resolution images
of the H$_2$ 1-0 S(1) line emission obtained with the Son of ISAAC (SOFI) at the New Technology Telescope (NTT)
of the Horsehead nebula. 
These observations are analysed in combination with H$\alpha$ line emission, aromatic dust, CO and dust continuum emissions. 
The Horsehead nebula illuminated by the O9.5V star $\sigma$ Ori
($\chi \sim$ 60) presents a typical photodissociation region (PDR) viewed nearly edge-on and offers an ideal opportunity to study the gas density structure of a PDR.\\
The H$_2$ fluorescent emission observations reveal 
 extremely sharp and bright filaments associated with the illuminated edge 
of the nebula which spatially coincides with the aromatic dust emission.
Analysis of the H$_2$ fluorescent emission, sensitive to both the far-UV radiation field and the gas density,
in conjunction with the aromatic dust and H$\alpha$ line emission,
brings new constraints on the illumination conditions and the gas density in the outer PDR region.
Furthermore, combination of this data with millimeter observations of CO and dust 
continuum emission allows us to trace the penetration of the far-UV radiation field into the cloud and probe
the gas density structure throughout the PDR.
From comparison with PDR model calculations, we 
find that i) the gas density follows a steep gradient at the cloud edge, 
with a scale length of 0.02 pc (or 10'') and $n_H\sim 10^4$ and $10^5$ cm$^{-3}$ in the H$_2$ emitting and inner cold molecular layers respectively,
and ii) this density gradient model is essentially a constant pressure model, with $P\sim$4 $10^6$ K cm$^{-3}$.
The constraints derived here on the gas density profile
are important for the study of physical and chemical processes
in PDRs and provide new insight into the evolution of interstellar clouds.
Also, this work shows the strong influence of the density structure on the PDR spatial stratification
and illustrates the use of different tracers to determine this density structure.
}

\authorrunning{Habart et al.}
\titlerunning{Density structure of the Horsehead nebula PDR}

\maketitle

\section{Introduction}

Photo-dissociation regions at the edge of molecular clouds 
illuminated by bright stars are unique targets for the study of the interaction of stellar ultraviolet radiation with dense gas. Such interactions ionize and dissociate molecules and heat the gas. This is important for the evolution of the cloud as well as for the star formation associated with the cloud.

The thermal and chemical structure of the PDRs depends mainly
on two parameters, namely, the intensity of the incident far-ultraviolet 
(FUV, 6$<h\nu<$13.6 eV) radiation field, characterised by a scaling factor $\chi$ \footnote{The
scaling factor $\chi$ represents here the intensity of the incident FUV radiation field
in units of the average interstellar radiation field defined by Draine \cite[]{draine78}. For a discussion of the definitions of mean interstellar radiation field used in the literature on PDRs see Appendix B of \cite{allen2004}.}, and the hydrogen nucleus gas density, $n_H$.
PDR models show that due to the attenuation of the incident FUV radiation field, the gas temperature and chemical composition are spatially stratified between the illuminated edge and the inner layer of the cloud.
Studying observationally this spatial structure allows one to derive the local physical conditions (i.e., $\chi$ and $n_H$) and infer the dominant processes in PDRs. 
Unfortunately for most known PDRs, it is difficult to establish the geometry and
to precisely measure the FUV penetration depths
and the gas density profile, which is a key input parameter of the PDR models.
\par\bigskip 
In this paper, we study the structure of the Horsehead nebula photodissociation region
located at the western edge of the molecular cloud L1630 illuminated by the O9.5V star $\sigma$ Ori. 
This nearby PDR (d$\sim$400 pc) viewed edge-on
offers an ideal opportunity of studying the structure of a PDR.
It represents one of the sharpest mid-infrared (IR) filament (width: 10'' or 0.02 pc) detected in our Galaxy by the ISO (Infrared Space Observatory) camera \cite[]{abergel2003}.
In order to resolve properly the PDR outer structure, we have 
obtained imaging observations of the H$_2$ 1-0 S(1) line emission at 2.12 $\mu$m using SOFI on the European Southern Observatory NTT telescope. This line emission is 
very sensitive to both the FUV radiation field and the gas density
and the angular resolution of this observations (1'') is $\sim$5-10 times better than previous 
infrared or molecular and atomic lines conducted in this region \cite[]{zhou93,kramer96,abergel2003,pound2003,teyssier2004}.
Thus, unprecedented constraints on the physical conditions at small scale of the illuminated material can be derived.
Also, in order to trace the PDR spatial stratification, we compare this near-IR data with the aromatic dust and H$\alpha$ line emission as well as
millimeter data of CO and dust continuum emission recently taken at the IRAM (Institut de Radio-Astronomie Millim\'etrique) 30m telescope \cite[]{abergel2003,teyssier2004} and the IRAM Plateau de Bure Interferometer \cite[PdBI,][]{pety2004}.
A detailed comparison of the observations with PDR model calculations
allows us to infer the gas density structure throughout the Horsehead nebula PDR.

The paper is organised as follows. In Sect. \ref{observations}, we present
the observations. The PDR models we used are described in Sect. \ref{model}.
The results of the calculations which explore the effect of varying gas density and of projection effects
are presented and compared to the observations in
Sect. \ref{uniform}, \ref{comparison_obs_mod}, \ref{gradient} and \ref{projection}.
In Sect. \ref{discussion}, we discuss our results.
Our paper conclusions are summarized in Sect. \ref{conclusion}.

\section{Observations}
\label{observations}

\begin{figure}[htbp]
\leavevmode
\begin{minipage}[l]{9cm}
\centerline{ \psfig{file=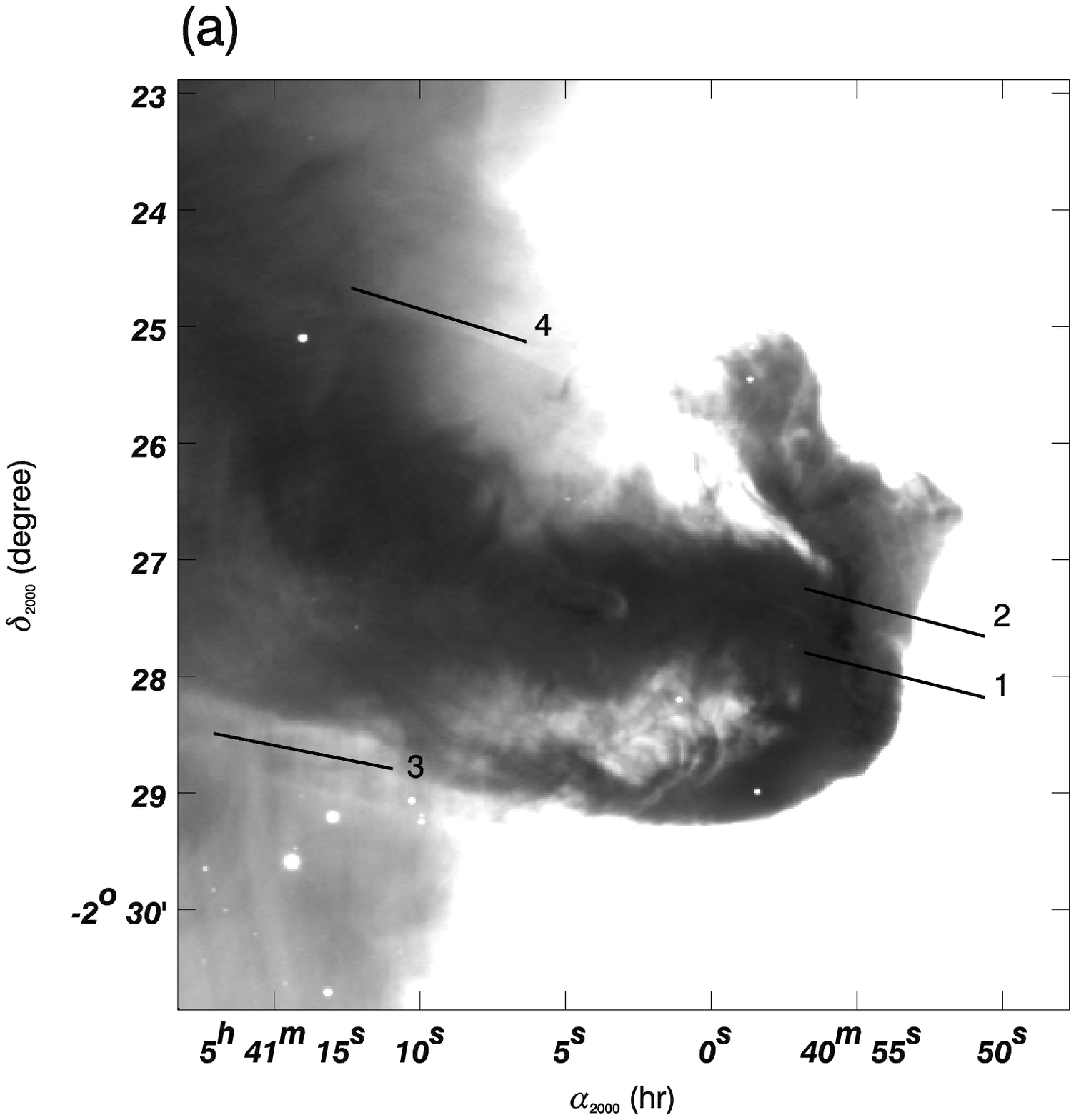,width=9cm,angle=0} }
\end{minipage}
\begin{minipage}[l]{9cm}
\centerline{ \psfig{file=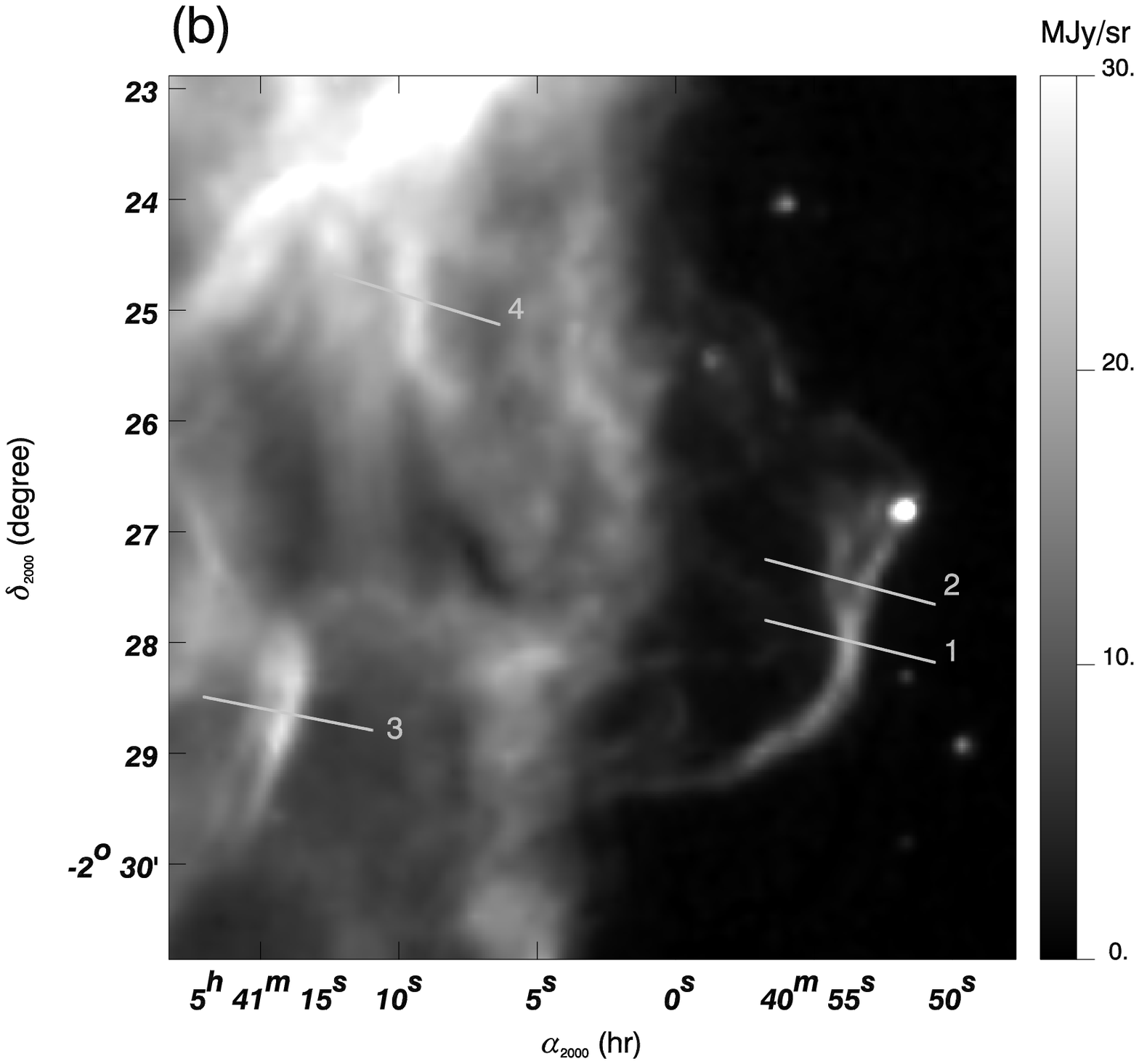,width=9cm,angle=0} }
\end{minipage}
\caption{\em (a) Map of the Horsehead nebula in H${\alpha}$ line emission obtained at the 0.9 m KPNO telescope 
(by Bo Reipurth and J. Bally). 
(b) Map in the LW2 filter (5-8.5 $\mu$m) of ISOCAM \cite[6'' of angular resolution,][]{abergel2003}. The brightness profiles along the cuts marked here are shown in Fig. \ref{cut_sofi}. The cuts are rotated by 14$^{\circ}$ to be in direction 
of the exciting $\sigma$ Ori star situated at $\alpha _{2000}$=05h38mn45.00s, $\delta _{2000}$=-02$^{\circ}$36'00''.
}
\label{map_Halpha}
\end{figure}

\subsection{Horsehead nebula}
\label{horsehead}

In the H$\alpha$ emission line, the Horsehead nebula (B33) emerges from the 
edge of L1630 as a dark cloud in the near side of the HII region IC 434 (see Fig. \ref{map_Halpha}a).
$\sigma$ Ori, a O9.5V system with $T_{eff}\sim$33,000 K, 
is in this region the main heating source \cite[]{abergel2003}.

Emissions of the dust and gas associated with the Horsehead nebula
have been mapped from the mid-IR to millimeter wavelengths \cite[]{abergel2003,pound2003,teyssier2004,pety2004}. 
In the ISO camera (ISOCAM) image taken in the LW2 filter (5-8.5 $\mu$m),
which is dominated by the emission of aromatic dust (hereafter referred to as PAHs or Polycyclic Aromatic Hydrocarbons),
the edge of the Horsehead nebula
is delineated by a bright and narrow filament (see Fig. \ref{map_Halpha}b).
The PAH emission, which is found to scale with the FUV radiation
field \cite[for $\chi$ ranging between 1 and $\sim 10^4$,][]{boulanger98b}, 
is a direct tracer of the local intensity of the incident radiation.  
Based on this, \cite{abergel2003} propose that this filament 
 is due to dense material illuminated edge-on by 
 $\sigma$ Ori, so that matter beyond the border should not be affected by the incident radiation field. Moreover, this structure must be seen nearly edge-on.  A large inclination ($\sim$30$^{\circ}$ or more)
relative to the line of sight would produce a broader filament. 
From the projected distance between the star and the interface,
$d_{proj}\sim$3.5 pc, and the spectral type of $\sigma$ Ori,
the FUV intensity of the incident radiation field
is $\chi =$60 (in units of the Draine field, 2.7 10$^{-3}$ erg s$^{-1}$ cm$^{-2}$).

The gas density derived from the penetration depth of the FUV radiation is a few times 10$^4$ cm$^{-3}$ \cite[]{abergel2003}.
This value is consistent with the analysis (using Large Velocity Gradients calculations)
of CO emission lines obtained at the IRAM 30-m telescope
 which suggests that beyond the edge the
gas density is of the order of (or superior) to 1-4 10$^4$ cm$^{-3}$ \cite[]{abergel2003}.

However, the angular resolution of ISOCAM (6'') is still
insufficient to resolve properly the density structure at the edge of the cloud.
Furthermore, the infrared emission of the aromatic dust depends on their abundance which can vary throughout the interface \cite[]{boulanger88b,boulanger90a,bernard93}.
In this context, we have obtained high resolution images
of the H$_2$ fluorescent emission which represents a powerful tool for 
studying the structure of the illuminated gas at the PDR edge.

\subsection{SOFI/NTT observations and data processing}
\label{sofi}

Using the SOFI/NTT infrared imager-spectrometer,
we have obtained images in the 1-0 S(1) emission line of H$_2$ at 2.12 $\mu$m and in the adjacent continuum filters over a section of the western edge of L1630. 
The image scale is 0.29'' per pixel, giving an observed field 
of $\sim$5'$\times$5'. 
We observed three fields : one centered on the Horsehead nebula ($\alpha _{2000}$ = 5h40mn56s, $\delta _{2000}$ = -2$^{\circ}$27'04.9'')
and two others taken respectively at 110''S 210''E 
 and 125''N 130''E 
 of the Horsehead nebula field. The Horsehead nebula 
field was observed in the two band filters H$_2$ 1-0 S(1) ($\lambda _c$=2.124  $\mu$m; $\Delta \lambda$=0.028 $\mu$m) and Br$\gamma$ ($\lambda _c$=2.167  $\mu$m; $\Delta \lambda$=0.028 $\mu$m) over two nights
in March 2000, while the two other fields were 
observed over two nights in January 2002 in the H$_2$ 1-0 S(1), NB ($\lambda _c$=2.09 $\mu$m; $\Delta \lambda$=0.02 $\mu$m) and Br$\gamma$ (only for one field)   
filters. 
The seeing was measured to be $\sim$1''.

Observation and data reduction were performed using standard procedures 
for near-IR imaging. A set of frames slightly offset in telescope position were obtained. 
They were then realigned and co-added to produce the final image.
A calibration was obtained with observations of standard stars from the
NICMOS list (N9115, N9116 and N9118). Calibration uncertainties are about 10\%.

\begin{figure*}[htbp]
\centerline{ \psfig{file=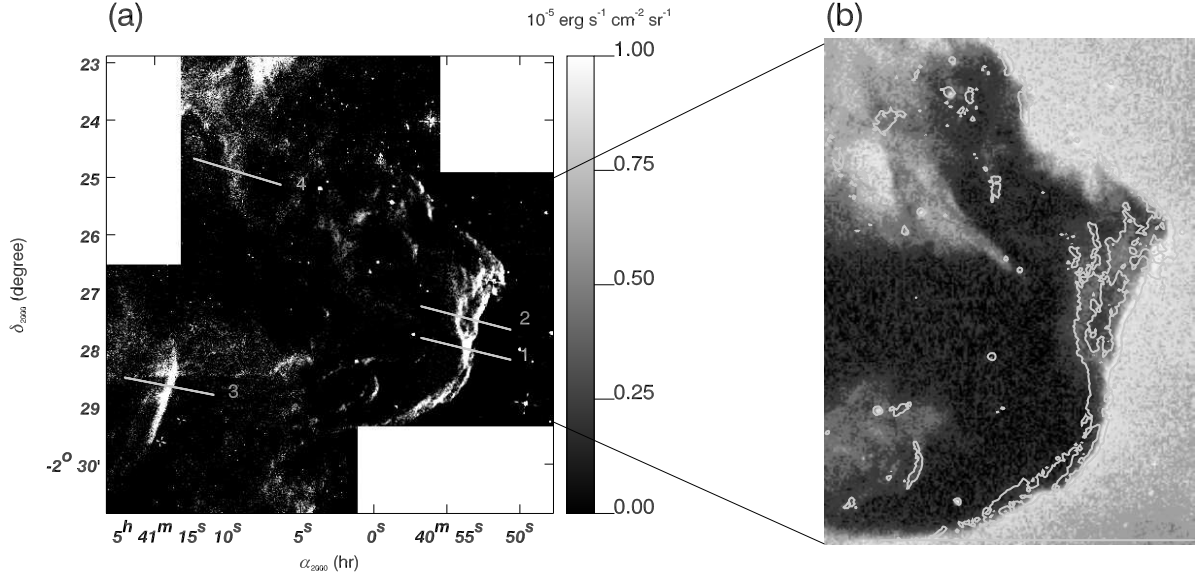,width=18cm,angle=0} }
\caption{(a) SOFI map in the 1-0 S(1) H$_2$ line emission 
 with the continuum subtracted. The seeing was about 1''.
The brightness profiles along the cuts marked here are shown in Fig. \ref{cut_sofi}.
(b) ESO-VLT composite colour image %($B$, $V$ and $R$ bands) 
with contour of the 1-0 S(1) H$_2$ line emission (level is 0.5 $10^{-5}$ erg s$^{-1}$ cm$^{-2}$ sr$^{-1}$). The seeing was about 0.75''.}
\label{map_sofi}
\end{figure*}

\begin{figure*}[htbp]
\centerline{ \psfig{file=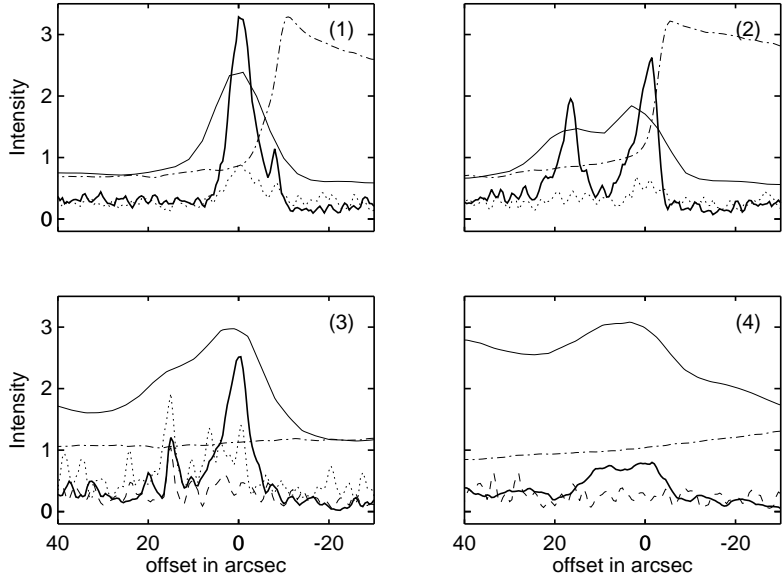,width=12cm,angle=0} }
\caption{Brightness profiles measured in the H$_2$ 1-0 S(1) filter (solid thick lines, $10^{-5}$ erg s$^{-1}$ cm$^{-2}$ sr$^{-1}$) and
in the adjacent continuum filters - Br$\gamma$ (dotted lines shown in cuts 1, 2 and 3) and NB (dashed lines, shown in cuts 3 and 4) - along the cuts shown in Fig. \ref{map_sofi}. 
We also show the ISOCAM emission in the LW2 filter 
(solid thin lines, MJy/sr $\times 0.1$) and the H$\alpha$ line emission (dotted-dashed lines, scaled to the H$_2$ emission of cut 1). The offset origins of the cuts correspond to the H$_2$ emission peaks.
}
\label{cut_sofi}
\end{figure*}

In Fig. \ref{map_sofi}, we show a composite map of the three fields observed
in the 1-0 S(1) H$_2$ line emission with the continuum subtracted. 
To substract from the continuum, we have used the NB filter
when available, and the Br$\gamma$ filter otherwise.
The NB filter is in fact narrower and less likely to be contamined by other gas lines.
In Fig. \ref{cut_sofi}, we show brightness profiles measured in 
the H$_2$ 1-0 S(1) filter as well as in the adjacent continuum filters towards 4 cuts
(shown in Figs. \ref{map_Halpha} and \ref{map_sofi}): the cut 1 goes through the main bright filament seen at the edge of the Horsehead nebula; the cut 2 through the two narrow filaments situated at 40''N; the cut 3 towards the other bright filament seen inside the L1630 cloud; and the cut 4 through a weak larger filament  situated inside the L1630 cloud.
Concerning the accuracy of our measured H$_2$ 1-0 S(1) line intensity,
we note that when using the Br$\gamma$ filter, the 1-0 S(1) H$_2$
line emission subtracted from the continuum can decrease by $\le$20\%.
Hence we assign an accuracy of about 20\% to the
measurements of the 1-0 S(1) line intensity.

\nopagebreak

\subsection{H$_2$ fluorescent, H$\alpha$ and PAH emissions}
\label{h2_halpha_lw2}

The SOFI data reveals extremely sharp filaments (width:$\sim$5'' or 0.01 pc) spatially coincident with the aromatic dust emission (see Figs. \ref{map_Halpha}, \ref{map_sofi} and \ref{cut_sofi}). 
Since the H$_2$ fluorescent emission is very sensitive 
to both the FUV radiation field and the gas density, 
this confirms that the mid-IR filament seen
by ISOCAM at the Horsehead nebula edge is due to illuminated dense material 
seen nearly edge-on.
Towards more tenous material, the H$_2$ emission would be weaker and more extended.
Also, a large inclination relative to the line of sight
would produce broader H$_2$ emission filaments.
Furthermore, as expected for a PDR seen edge-on, we find that at the edge of the nebula,
the H$_2$ fluorescent emission
peak is displaced further into the cloud than the H$\alpha$ line 
emission which traces the ionization front (see Fig. \ref{cut_sofi} panel 1).
Nevertheless, we note that the H$\alpha$ line emission decreases going towards the cloud by only a factor of 5 (see Fig. \ref{cut_sofi} panel 1). This indicates that $\sigma$ Ori is probably located slightly closer to us
than the Horsehead (if $\sigma$ Ori were beyond the Horsehead, the H$\alpha$ line emission towards the cloud would diminish by a much large factor). Moreover, it must be emphasized here that since the H$\alpha$ line emission 
is a strong function of the extinction along the line of sight, 
it is impossible to make qualitative predictions.
In cases such as the cuts 3 and 4 taken towards cloud,
the H$\alpha$ line emission profile 
probably has nothing to do with the appearance of the PDR (see below).

In the SOFI data, we also discover sub-structures unresolved by ISOCAM. 
At the edge of the Horsehead nebula, we can in particular clearly distinguish two narrow filaments well separated at several places along the interface (see Figs. \ref{map_sofi} and \ref{cut_sofi})
 except around $\delta _{2000}$=-2$^{\circ}$28'00'' where they are likely superposed. This filamentary structure is also
 seen in the visible (see Fig. \ref{map_sofi}b) presumably due to the scattered light.
We believe that these infrared filaments represent the PDR edge 
at different positions along the line of sight.

Inside the L1630 cloud, the bright filament seen in cut 3 may also represent illuminated dense material seen edge-on. 
In fact, the H$_2$ emission is shown to delineate a narrow 
filament spatially coincident with the aromatic dust emission.
Along this filament, we note that the H$\alpha$ intensity is roughly constant and does not present any front
as at the edge of the Horsehead nebula.
It is possible that this filament structure is seen beyond the Horshead nebula.
The structure seen in cut 4 may be 
on the other hand due to more tenous material since the H$_2$ emission in that case
is weaker and more extended. 
Also, we note that the H$_2$ 1-0 S(1) line to LW2 emission ratio towards cut 3 is
roughly similar to that measured in the cuts 1 and 2 
while it is lower for the cut 4 by a factor of 2-3 (see Fig. \ref{cut_sofi}).
This may reflect differing local gas densities. The dependence of the H$_2$ to aromatic
dust emission ratio with gas density
is discussed in Sect. \ref{model_obs}.

In the next section, in order to study the density structure throughout the Horsehead nebula 
PDR, we combine this data with recent millimeter wave observations.

\begin{figure}[htbp]
\leavevmode
\begin{minipage}[l]{9cm}
\centerline{ \psfig{file=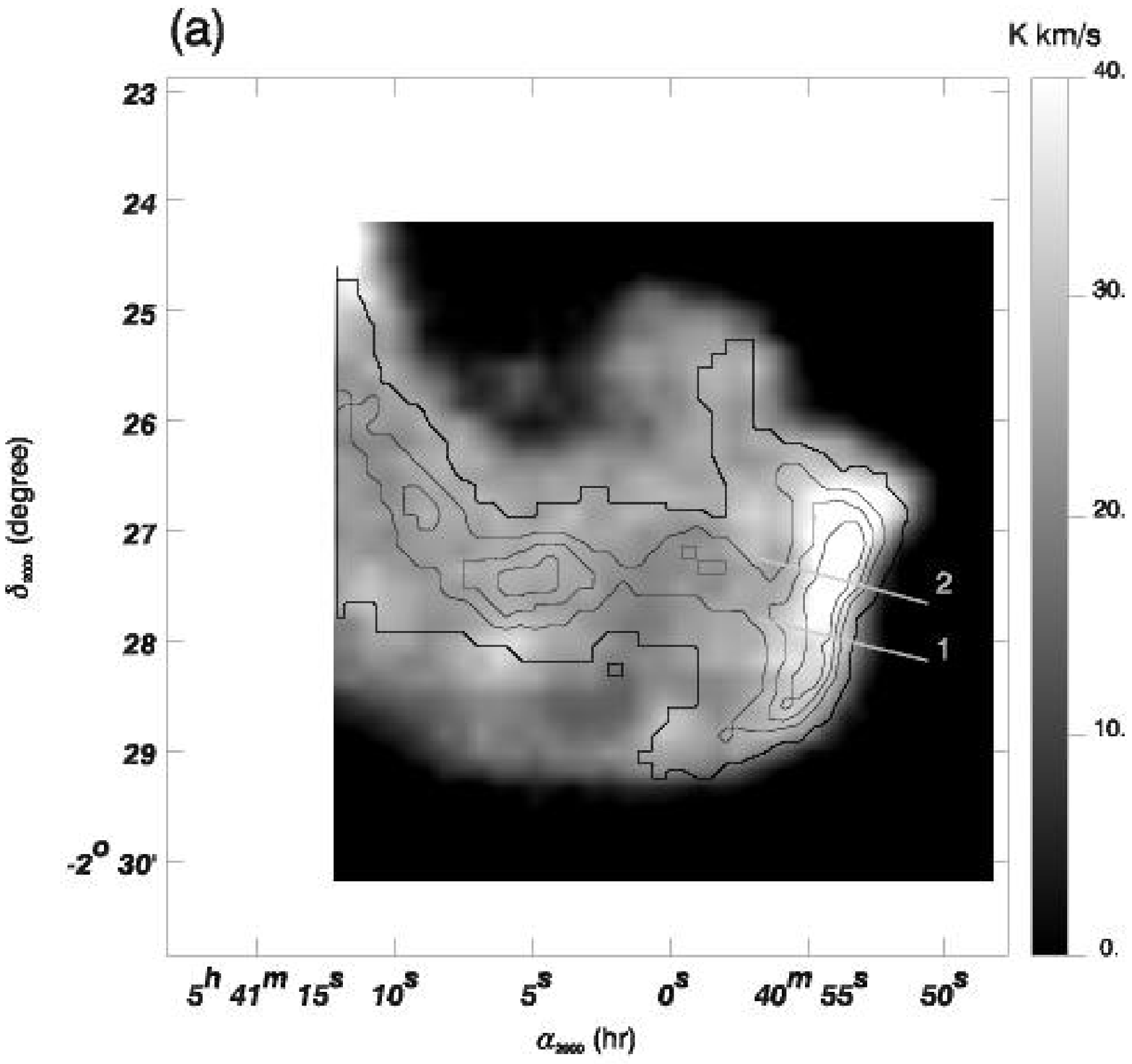,width=9cm,angle=0} }
\end{minipage}
\begin{minipage}[c]{9cm}
\centerline{ \psfig{file=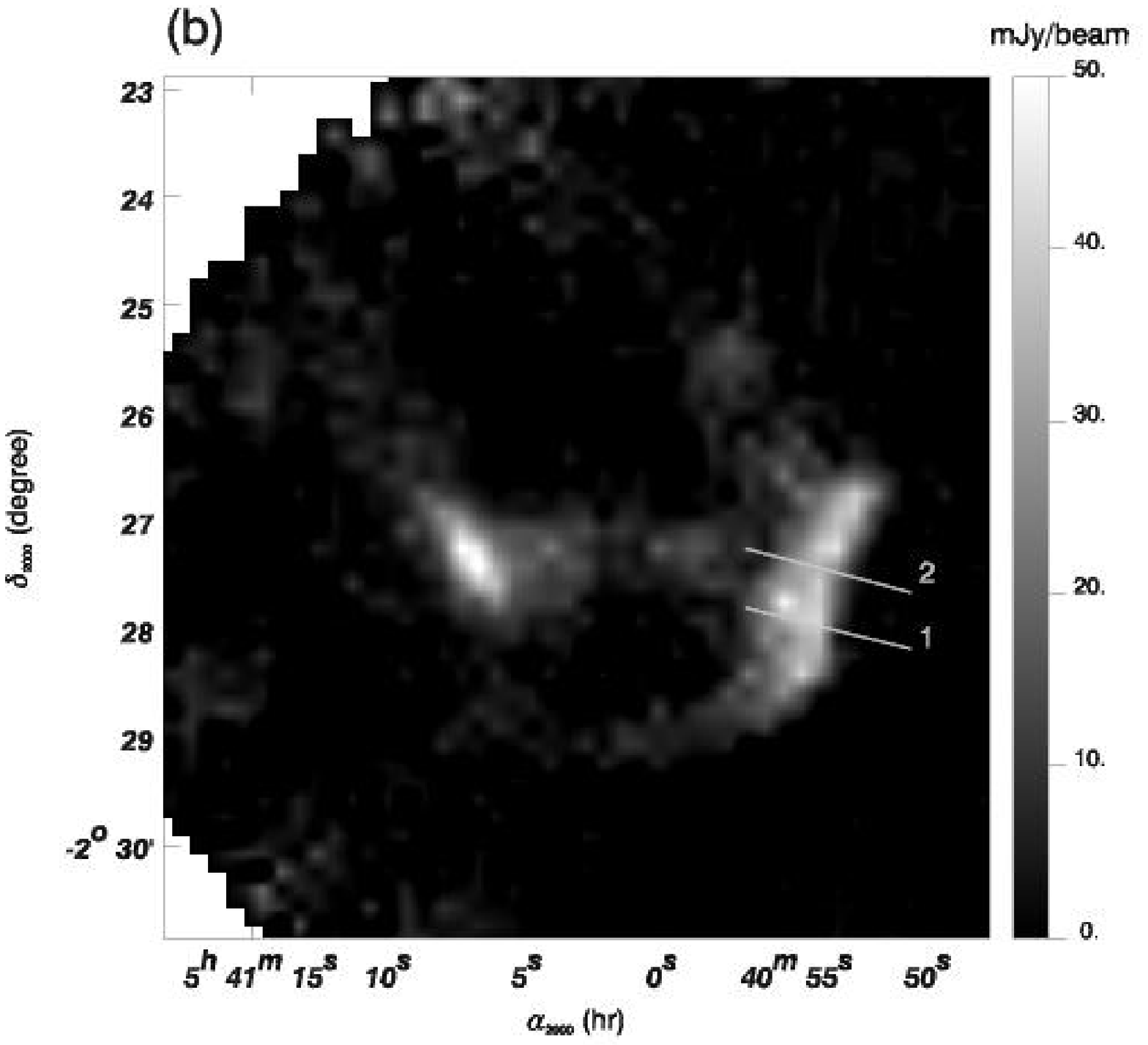,width=9cm,angle=0} }
\end{minipage}
\caption{\em (a) Integrated intensity map of the $^{12}$CO(3-2) line emission made with the CSO 
(grey scale) with contours of the C$^{18}$O(2-1) line emission (levels are 4.5, 6, 7.5, 9 K km/s) made
with the IRAM-30m telescope. (b) Map of the 1.2 mm dust continuum emission.
The brightness profile along the cut 1 is shown in Fig. \ref{cut_proj}.}
\label{map_mm_proj}
\end{figure}

\begin{figure}[htbp]
\begin{center}
\leavevmode
\centerline{ \psfig{file=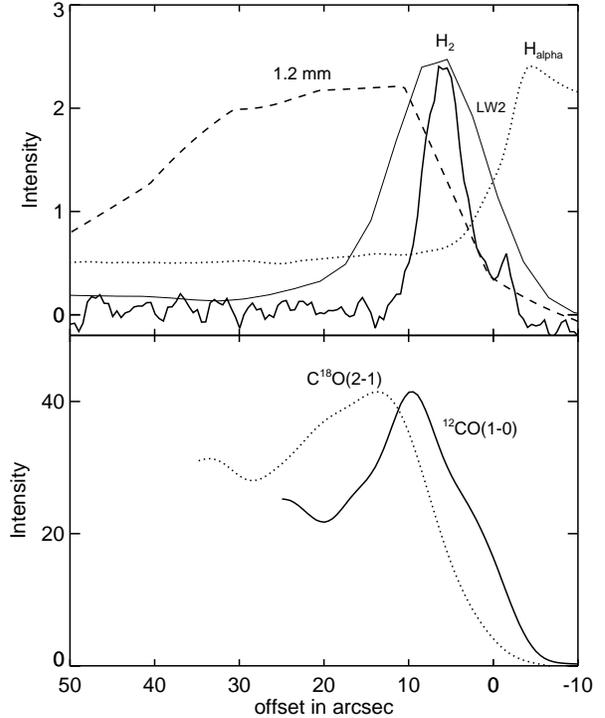,width=9cm,angle=0} }
\end{center}
\caption{\em Emission profiles throughout the edge of the Horsehead nebula along the cut 1 (shown in Figs. \ref{map_Halpha}, \ref{map_sofi} and \ref{map_mm_proj}).
Upper panel: H$\alpha$ line emission (dotted lines) scaled to the H$_2$ 1-0 S(1) line emission (solid thick lines, 10$^{-5}$ erg s$^{-1}$ cm$^{-2}$ sr$^{-1}$); ISOCAM emission in the LW2 filter (solid thin line, $MJy/sr$ $\times 0.14$) substracted from the background emission taken to be the average of the minimum
value of this extending cut (6.2 MJy/sr); dust continuum emission at 1.2 mm (dashed line, mJy/beam$\times$0.05).  
Lower panel: Emission of the $^{12}CO$ (1-0) (solid lines, K km/s) and $C^{18}O$(2-1) (dotted lines, K km/s$\times$5) 
lines obtained from with the PdBI. The origin offset is taken where the H$\alpha$ intensity is decreased by 50\%.}
\label{cut_proj}
\end{figure}

\subsection{Combination with millimeter data of CO and dust continuum emission}
\label{millimeter}

Recently, \cite{pety2004} have obtained high spatial resolution maps of
CO and isotopomers emission lines, namely $^{12}$CO(1-0), $^{12}$CO(2-1) and C$^{18}$O(2-1)
 at the Plateau de Bure Interferometer (PdBI). 
These data were obtained with typical resolutions of 5'' and 2.5'' at 2.6 mm and 1.3 mm respectively for $^{12}$CO and of 6'' at 1.4 mm for C$^{18}$O.
We used these data to trace the spatial distribution of the molecular gas beyond the edge.

Secondly, in order to measure the intensity ratios of several CO emission lines, 
we used maps of the ($J=1-0$) and ($J=2-1$) transitions of $^{12}$CO, $^{13}$CO and C$^{18}$O 
obtained at the IRAM 30-m telescope by \cite{abergel2003,teyssier2004}.
 These maps have an angular resolutions of 10.5'' and 22'' at 1.3 mm and 2.6 mm respectively. 
We also use the map of the ($J=3-2$) transition of CO and $^{13}$CO obtained at the CSO (M. Gerin, private communication).
In Fig. \ref{map_mm_proj}(a), we show the integrated intensity map of the $^{12}$CO(3-2) line
emission with contours of the C$^{18}$O(2-1) line emission.
The comparison between these two maps show particularly 
the effect of optical depth in the molecular cloud. 
The emission of the rare and  probably optically thin isotopomer C$^{18}$O probe
the inner shielded regions, while the $^{12}$CO optically thick emission
likely comes from outer layers.
The $^{12}$CO optically thick emission could trace the front face-on side
of the Horsehead nebula.
Then, in the following, in order to probe the structure in the inner shielded
regions, we will focus more on the analysis
on the optically thin C$^{18}$O lines emission.

Finally, in order to trace the quantity of matter towards the PDR, we used a map of the 1.2 mm dust continuum emission taken with MAMBO at the IRAM 30-m telescope \cite[]{teyssier2004}.
This map shown in Fig. \ref{map_mm_proj}(b) has a resolution of $\sim$11''.
We note that the C$^{18}$O emission
 coincides generally well with the emission of the dust at 1.2 mm,
at least in the PDR region where no depletion is expected.
That confirms that the C$^{18}$O probes the inner denser regions and that the emission of this isotopomer is optically thin.

In Fig. \ref{cut_proj}, we show the brightness profiles of 
different tracers across the PDR (cut 1).
A layered appearance is clear; moving away from the excitation source,
 the H${\alpha}$, H$_2$, $^{12}$CO and C$^{18}$O lines emissions appear in succession.
This spatial stratification is typical for an edge-on PDR.
In front of the PDR ionization front (as traced by the H${\alpha}$ line emission), the H$_2$ and CO molecules 
are photodissociated; at optical depth of about  $A_V\lesssim 1$,
the dissociating photons have been sufficiently attenuated to cause neutral H$^0$
to be  transformed into H$_2$; similarly, at $A_V\sim 2-4$,  the carbon balance shifts from C$^+$ to neutral C$^0$ and CO.
 Hence, the separation between the  H$_2$ and CO emission peaks measures directly the FUV penetration size scale. 
From Fig. \ref{cut_proj}, we estimate that the observed scale size of the region is about $\sim$5-10'' per magnitude of visual extinction. Further, adopting a hydrogen column density, $N_H$, per magnitude of visual extinction, $N_H/A_V=1.85~10^{21}$ cm$^{-2}$ and a distance of 400 pc, we derive an average gas density of $\sim$5 10$^4$ cm$^{-3}$ for a homogeneous region viewed edge-on. This value is comparable to the previous gas density derived by \cite{abergel2003}.

In the following, in order to analyse in details the observed spatial stratification and constrain the gas density 
structure, we compare the observations with PDR model calculations.

\section{Comparison with PDR models}
\label{model_obs}

\subsection{PDR models}
\label{model}

We use an updated version of the stationary PDR model described in \cite{lebourlot93} \cite[for a recent paper see][]{lepetit2002}.
In this model, a PDR is represented by a
semi-infinite plane-parallel slab with an incident radiation field.
The input parameters are {\rm (i)} the incident FUV field $\chi$, and {\rm (ii)} the hydrogen nucleus
gas density $n_H$. 
In our models, in addition to the ever existing isotropic interstellar radiation field ($\chi$=1),
we consider a blackbody incident spectrum with $T_{eff}=33,000$ K
and an FUV radiation strength of $\chi$=60 (in units of Draine field, see Sect. \ref{horsehead}).
This in principle is an upper limit and hence the effect of changing the value of $\chi$ is discussed in Sect. \ref{discussion}.
For the gas density, we will consider constant and gradient density models as well as
isobaric models. 
With these inputs, the model solves the chemical and thermal balance starting from the slab edge at each $A_V$-step in the cloud.
\par\bigskip
The dominant process governing the chemical balance is
 the photodissociation of H$_2$, the rate
of which is determined by both H$_2$ self-shielding and attenuation by
 dust. For the dust extinction properties, we use the analytical fit of 
\cite{fitzpatrick88} to Galactic average extinction curve.
The radiative transfer in the absorption lines of H$_2$ and CO is treated 
in detail and the individual line profiles are described with the prescription of \cite{Federman79}.
We adopt an H$_2$ formation rate $R_f$ (cm$^3$ s$^{-1}$) constant throughout the PDR and equal to
the standard value inferred from observations of H$_2$ UV absorption lines in interstellar diffuse clouds, i.e. $R_f$=3 10$^{-17}$ cm$^3$ s$^{-1}$ \cite[]{jura75}.
This assumption is discussed in Sect. \ref{discussion}.

The chemical network involves about 130 chemical species and 1100
reactions. For the elemental abundances, we take the following {\it standard} values : He/H=0.1, C/H = 1.35 10$^{-4}$,
O/H = 3.19 10$^{-4}$, N/H=7.5 10$^{-5}$, S/H=1.85 10$^{-6}$, Fe/H=1.5 10$^{-8}$. As $^{13}$CO and C$^{18}$O observations are reported, we have added to the reaction set, the main isotopic molecules involving $^{13}$C (with $^{13}$C/H=1.6 10$^{-6}$) and $^{18}$O  (with $^{18}$O/H = 6 10$^{-7}$) and
introduced the corresponding fractionation reactions \cite[]{graedel82}.
Also, the photodissociation of the CO isotopomers is treated in details.

For the physical conditions prevailing in the Horsehead nebula ($\chi = 60$, 
$n_H \ge$ $10^4$ cm$^{-3}$, see Sect. \ref{horsehead}), 
the thermal balance is mainly determined by the photoelectric effect
on small dust grains. The photoelectric heating rate is derived 
from the formalism of \cite{bakes94}. 

\par\bigskip
The gas line emission profiles are calculated assuming a  plane-parallel edge-on geometry.
The gas line intensity (in erg s$^{-1}$ cm$^{-2}$ sr$^{-1}$) integrated along a given line of sight $i$ can be written as:
\begin{equation}
I_{\nu} (i) = \frac{J_{\nu}(i) \times \beta (i)}{4\pi} \times l_{PDR} 
\label{Eq:intensity}
\end{equation}
%$i$ labels the $A_V$-layers from the model along the line of sight;
with $J_{\nu}(i)$ the line emissivity (in erg s$^{-1}$ cm$^{-3}$) extracted from the plane-parallel model at the $A_V (i)$-layers, $\beta$ the escape probability %from layer $i$ to the cloud edge
 along the line of sight and $l_{PDR}$ (in cm) the length of the PDR along 
the line of sight. 
For $\beta$, we use the formalism of \cite{tielens85} 
in their Appendix B with a turbulent Doppler width $\delta v_d$= 2 km s$^{-1}$. % (larger observed in the CO lines, see Sect. \ref{millimeter}).
This latter assumption is only important for the $^{12}$CO (1-0), (2-1) and (3-2) lines
which do not greatly affect our conclusions. 
Optically thin lines (such as H$_2$ and C$^{18}$O) are unaffected by the assumption used to define $\beta$.

We compute the aromatic dust emission similarly by :
\begin{equation}
I_{PAH} (i) = \frac{J_{PAH}(i)}{4\pi} \times l_{PDR}
\end{equation}
with $J_{PAH}$ the power emitted by PAHs computed with the PAH absorption cross-section of \cite{verstraete92}. 
We assume the optical properties and the abundance of aromatic dust to be constant across the PDR.
The carbon locked up in PAHs ($a \lesssim$10 \AA) has an abundance of $[C/H]_{PAH} \simeq 5$ 10$^{-5}$ inferred from the 12 $\mu$m emission per hydrogen in typical Galactic cirrus 
 \cite[]{boulanger88} and PDR \cite[]{habart2001a} and from comparison between observations of dust galactic emission and extinction with detailed model
calculations \cite[]{desert90,dwek97,li2001a}.

We compute the 1.2 mm dust continuum emission in the Rayleigh-Jeans approximation of the Planck function as \cite[]{motte98} : 
\begin{equation}
\frac{S_{\nu} (i)}{10~\rm{mJy/beam}}  =  \frac{N_{H_2}(i)}{4~10^{21}~\rm{cm}^{-2}} \times \frac{T_{dust}(i)}{20~\rm{K}} \times \frac{k_{\nu}}{0.005~\rm{cm}^2\rm{g}^{-1}}
\label{Eq.submm}
\end{equation}
with $N_{H_2}(i)=n_{H_2}(i) \times l_{PDR}$ the column density of H$_2$ along the line of sight $i$, $T_{dust} (i)$ the dust temperature and $k_{\nu}$ the dust opacity per unit mass. The thermal dust temperature is calculated
 using the analytic expression of \cite{hollenbach91}
and we adopt $k_{\nu}$ constant throughout the PDRs and equal to 0.0035 $\rm{cm}^2~\rm{g}^{-1}$  \cite[as recently measured in dense
clouds,][]{bianchi2003}.
This assumption is discussed in Sect. \ref{discussion}.

To calculate the gas and dust emissions, we assume that $l_{PDR}$ (the length of the
PDR along the line of sight) is constant throughout the PDR.
Note that this assumption is supported by the fact that H$_2$ and PAHs emissions are only seen at the
edge of the nebula, so that matter beyond the edge is not affected
by the incident radiation field. A round shape would produce H$_2$ or PAH emission further into the Horsehead nebula.

\subsection{Model predictions for a uniform gas density}
\label{uniform}

As a first step, we consider models with a uniform gas density with $n_H=$2 10$^4$ - 2 10$^5$ cm$^{-3}$ in the range expected  in the Horsehead nebula PDR (see Sects. 
\ref{horsehead} and \ref{millimeter}).

\begin{figure}[htbp]
\leavevmode
\begin{minipage}[l]{7cm}
\centerline{ \psfig{file=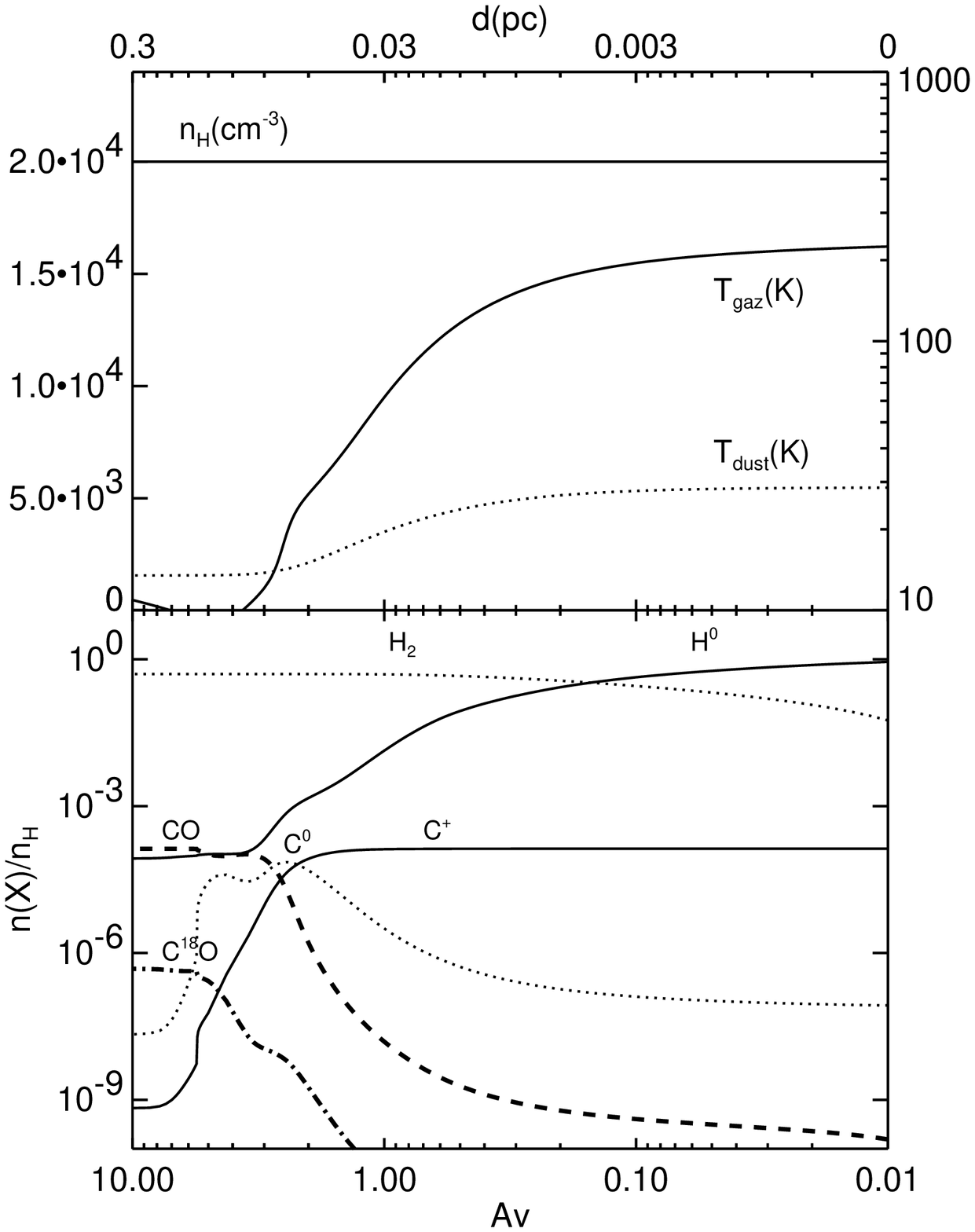,width=7cm,angle=0} }
\end{minipage}
\begin{minipage}[l]{7cm}
\centerline{ \psfig{file=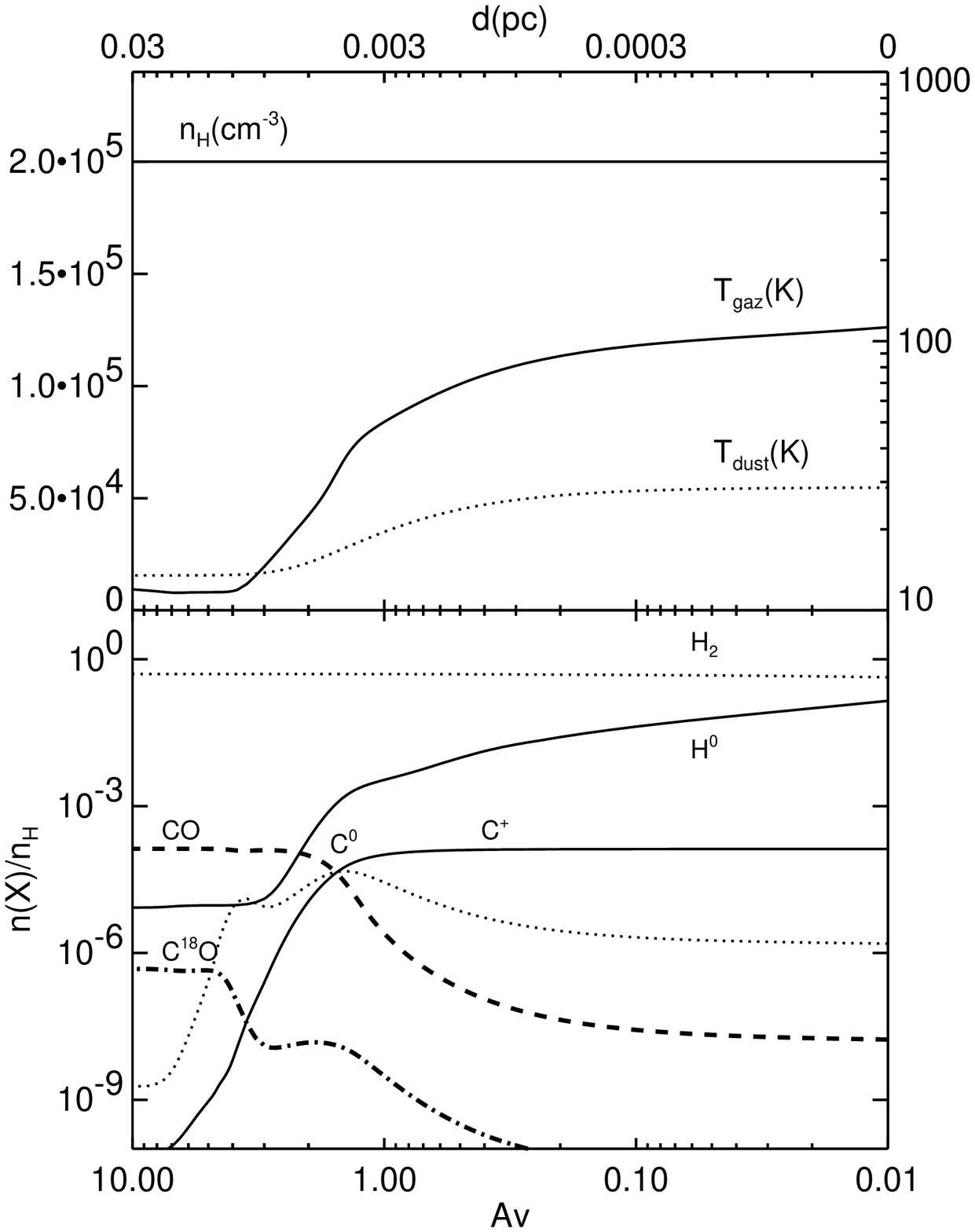,width=7cm,angle=0} }
\end{minipage}
\caption{\em Model predictions for $\chi$ = 60 and $n_{H}=2~10^4$ cm$^{-3}$ (upper panel) and $n_{H}=2~10^5$ cm$^{-3}$ (lower panel).
The visual extinction $A_V$ is taken from the PDR outer edge.
The scale on the upper axis shows the corresponding distance (in pc) from the PDR outer edge.
Upper panels: Gas and dust temperature. Lower panels: Abundances relative
to total hydrogen. The chemical PDR structure is characterized by the H$^0$/H$_2$ and C$^+$/C$^0$/CO transitions.}
\label{param1}
\end{figure}

In Fig. \ref{param1}, we show the model predictions of the thermal and chemical stratification as a function of depth into the PDR.
The chemical structure is characterized by the H$^0$/H$_2$ transition occuring
 at $A_V\sim 0.1$ and 0.01 for $n_H=$2~10$^4$ and 2 10$^5$  cm$^{-3}$ respectively, 
and the C$^+$/C$^0$/CO transition occuring at  $A_V\sim 1-4$ for both values of $n_H$. This can be understood as following.
For the low $\chi$/$n_H$ ratio adopted here ($\sim$0.01-0.001),
 the H$^0$/H$_2$ transition is mainly driven by the self-shielding in the H$_2$ lines.
Also, for higher gas density H$_2$ self shields more efficiently, so that
H$^0$ transforms into H$_2$ closer to the edge.
On the contrary, due to the low abundance of CO in the outer region, 
the  C$^+$/C$^0$/CO transition is governed by the dust opacity and 
occurs at the similar optical depth for both the values of $n_H$.
Fig. \ref{param1} shows also that, as for the chemistry, the thermal structure is governed by the penetrating FUV photons.
Until $A_V \sim 1$, the decrease of the gas temperature reflects the dependence of the photoelectric heating with the FUV radiation field
intensity. Then, in the inner region  opaque to the FUV field, the gas is heated by gas-grains collisions, exothermal chemical reactions and cosmic rays, and so that the gas temperature is roughly constant.

\nopagebreak
\begin{figure}[htbp]
\leavevmode
\centerline{ \psfig{file=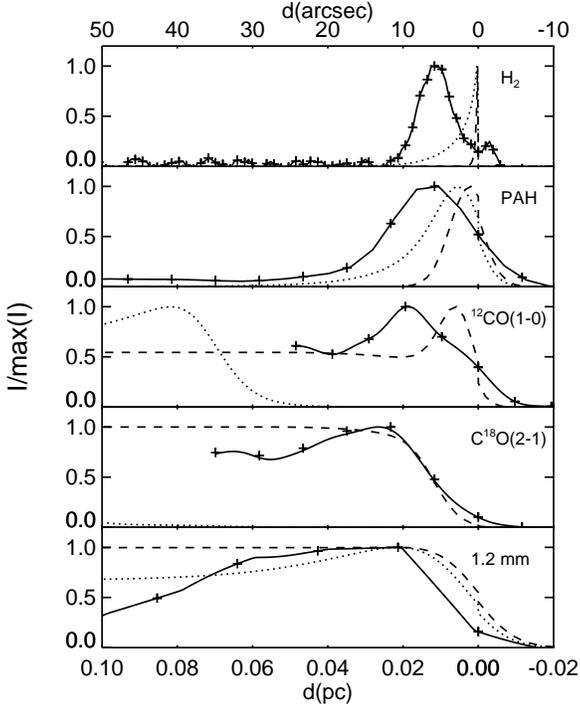,width=10cm,angle=0} }
\caption{\em  
Normalised emission profiles predicted by the PDR model with a constant gas density equal
to $n_H=2~10^4$ cm$^{-3}$  (dotted lines) and $n_H=2~10^5$ cm$^{-3}$  (dashed lines), 
and observed along the cut 1 
(solid lines with crosses), as a function of the distance from the PDR outer edge, $d$. The scale on the upper axis shows the corresponding offset in arcsec.
The predicted emission profiles have been convolved
with gaussian beam profiles of FWHM corresponding to the angular
resolutions of each observations (given in Sect. \ref{observations}).
For the observations, $d=0$ is taken where the H$\alpha$ intensity is decreased by 50\%.
}
\label{comp_obs_mod}
\end{figure}

In Fig. \ref{comp_obs_mod}, we show the 
normalised emission profiles predicted by the PDR model as a function of the distance from the PDR outer edge.  
As expected, the emission profiles of both the H$_2$ and CO lines depends strongly on the gas density. For higher $n_H$ the emission lines are narrower and peaks nearer the PDR edge.
In Table \ref{Table}, we give   at the emission peak
the intensity of H$_2$ and CO lines as well as several intensity ratios.
To estimate $l_{PDR}$ (the length of the PDR along the line of sight)
we use the H$_2$ column density derived from the 
observed 1.2 mm dust continuum emission peak (at $d\sim0.02$ pc or 10'')
and adopting the dust temperature predicted by the PDR model \footnote{To derive $l_{PDR}$ we do not use the H$_2$ fluorescent emission data since this emission could depend on projection effects due to sub-structures at the edge of the nebula as shown in Fig. \ref{map_sofi} and discussed in Sect. \ref{projection}.}.  
For $n_{H}=2~10^4$ cm$^{-3}$, 
we find that $T_{dust}\sim$22 K at $d\sim0.02$ pc (see Fig. \ref{param1})
and using Eq. \ref{Eq.submm} that $N_{H_2}\sim 1.5 ~10^{22}$ cm$^{-2}$.
For $n_{H}=2~10^5$ cm$^{-3}$, $T_{dust}\sim$13.5 K at $d\sim0.02$ pc (see Fig. \ref{param1})
and $N_{H_2}\sim 3 ~10^{22}$ cm$^{-2}$.
This implies $l_{PDR}\sim 0.5$ pc (or 5') for $n_{H}=2~10^4$ and 0.1 pc (or 1') for $2~10^5$ cm$^{-3}$, respectively. The value of  $l_{PDR}$ derived here for $n_{H}=2~10^4$ cm$^{-3}$
 which corresponds to about 2 times the extent of the filaments in the plane
of the sky (0.24-0.34 pc or 2-3') seems unlikely.
If we assume that $l_{PDR}$ is at most equal to the extent of the filaments in the plane of the sky, 
we conclude that $n_H$  is greater than $3~10^4$ cm$^{-3}$. 
However, this estimate is relevant for the gas into the cloud 
 and $n_H$ can be lower in the H$_2$ emission zone. 
For the sake of simplicity, we adopt for the model with $n_{H}=2~10^4$ cm$^{-3}$ 
the same  PDR length as for $n_{H}=2~10^5$ cm$^{-3}$, i.e., $l_{PDR}=0.1$ pc.

Table \ref{Table} shows first that the absolute intensity
of the  H$_2$ fluorescent line depends strongly on the gas density. 
The H$_2$ 1-0 S(1) line emissivity peak
 is two order of magnitude higher for $n_H=$2 10$^5$ cm$^{-3}$ than for $n_H=$2 10$^4$ cm$^{-3}$ . This can be understood as follows. 
At equilibrium between formation and dissociation of H$_2$, we have
\begin{equation}
R_f~n_H~n_{H^0} = R_{d}(0) \times \chi e^{-\tau _d} \times f_{s}(N(H_2)) \times n_{H_2}
\label{Eq_Rf}
\end{equation}
where $R_f$ is the H$_2$ formation rate, $R_{d}(0)\sim 0.1~R_a (0)$ 
the unshielded photodissociation rate per H$_2$ for $\chi$=1 
and with $R_a(0)\sim 5\times 10^{-10}$ s$^{-1}$ the photoabsorption rate, $e^{- \tau _{d}}$ and $f_{s}(N(H_2))$ the dust extinction and the H$_2$ self-shielding factors, respectively.
Since the intensity of the 1-0 S(1) line $I^f_{H_2}$ is proportional to the photoabsorption rate, we have $I^f_{H_2} \propto R_f n_H N(H^0)$ where $N(H^0)$ is the column density of atomic H atoms. Then, for an edge-on PDR we have $N(H^0) = n_{H^0} \times l_{PDR}$ and as $n(H^0)\sim n_H$ at the PDR edge, we thus have $I^f_{H_2} \propto n_H^2$. 
Moreover, we see that the H$_2$/PAH intensity ratio scales with the local gas density.
The emission of aromatic dust $I_{PAH}$ goes roughly as $\chi ~e^{- \tau _{d}} \times N_H \times [C/H]_{PAH}$ with $[C/H]_{PAH}$ the abundance of carbon locked up in PAHs. 
Thus, for an edge-on PDR $I_{PAH}$ scales with $n_H$ and $I^f_{H_2}/I_{PAH} \propto n_H$.

The CO line emission increases also with $n_H$; the optically thin lines of C$^{18}$O 
 are about 5 times higher for $n_H=$2 10$^5$ cm$^{-3}$ while the optically thick lines of $^{12}$CO
are about 2 times higher. The C$^{18}$O/$^{12}$CO (2-1) line intensity ratio which is mainly dependent on
the total column density is 3 times lower for $n_H=$2 10$^4$ cm$^{-3}$.
Finally, we find that the CO(2-1)/(1-0) and (3-2)/(1-0) 
intensity ratios are slightly higher for $n_H$=2 10$^5$ cm$^{-3}$.
This is probably due to the higher collision excitation
 and the slight increase of $T_{gas}$ in the CO emission region 
for $n_H$=2 10$^5$ cm$^{-3}$.

\begin{table*} \caption{Intensities and intensity ratios of H$_2$ and CO lines at the emission peak.}
\label{Table}
\begin{tabular}{llllllll}
\hline
\hline
                    & model 1 & model 2 & model 3 &model 4 & model 5 &  Observations\\
\hline
$d_{peak}$ (pc) $^a$  &&&&&&& \\
%&& \\
H$_2$          & 5 10$^{-2}$       & 0            & 0.003     &0.012            &0             &0.01\\
$^{12}$CO        &   0.08          & 0.005        & 0.016     & 0.022        &0.01            &0.02\\
C$^{18}$O      &  0.2            & 0.025         & 0.03     & 0.035          & 0.025           & 0.023    \\       
&& \\
Intensity (erg s$^{-1}$ cm$^{-2}$ sr$^{-1}$) $^b$  &&&&&&& \\
%&& \\
H$_2$ 1-0 S(1)  & 1 10$^{-4}$     & 1 10$^{-2}$  &1 10$^{-4}$&6 10$^{-5}$ &4 10$^{-5}$     & 2.4$\pm$0.5 10$^{-5}$\\ %(0.5)
H$_2$ 1-0 S(1)/PAH &  0.01          & 0.1         & 0.01      & 0.008     &0.005             & 0.004$\pm$0.0015 $^c$\\ %(0.001)
&& \\
$^{12}$CO(2-1) &   4 10$^{-7}$   & 8 10$^{-7}$  &7 10$^{-7}$&6 10$^{-7}$& 2 10$^{-6}$     & 8$\pm$0.05 10$^{-7}$ \\ %(0.04)
(2-1)/(1-0)         &   8             &  12          & 12        &   12      &   12                 & 13$\pm$1 \\%(0.01)
(3-2)/(1-0)         &   20            &  30          & 30        &   30      &   30                 & 20$\pm$2 \\%(0.1)
&& \\
$^{13}$CO(2-1)  &  7 10$^{-8}$  &2 10$^{-7}$ &2 10$^{-7}$&2 10$^{-7}$&2.8 10$^{-7}$ & 3$\pm$0.06 10$^{-7}$ \\%(0.06)
(2-1)/(1-0)         &  7              &  8           & 8            &   8    &   10                 &  11$\pm$2  \\ %(0.2)
(3-2)/(1-0)         &  15             &  20          & 20          &   20      &   20                &  13$\pm$0.7 \\%(0.65)
&& \\
C$^{18}$O(2-1)  & 2.6  10$^{-8}$    &  1.4 10$^{-7}$ &1.4 10$^{-7}$&1.2 10$^{-7}$&1.5 10$^{-7}$      &     1$\pm$0.05 10$^{-7}$      \\  %$^e$%(0.07)
(2-1)/(1-0)         & 9               &  10          & 10        &   10      &   10                  &   20$\pm$2    \\ %$^f$%(1.4)
\hline
\end{tabular}

Models 1 and 2 with an uniform gas density equal to $n_H=2~10^4$ and $2~10^5$ cm$^{-3}$, respectively.\\ 
Models 3 and 4 with a gas density gradient with $\beta=1$ and 4, respectively (see Sect. \ref{gradient}).\\
Model 5 with a gas density gradient with $\beta=1$ and taking into account inclination effect with $\theta=6^{\circ}$ (see Sect. \ref{projection}).\\
$^a$ Distance of the emission peak from the PDR outer edge. \\
$^b$ Intensity  at the emission peaks. 
$^c$ From the brightness in the LW2 filter (5-8.5 $\mu$m), which is dominated by the aromatic dust emission \cite[]{boulanger98}, we have estimated the aromatic dust emission using the following relationship : $I_{PAH}(2-15 \mu m) \simeq 2 \times \nu I_{\nu}(5-8.5 \mu m)$ based on ISOCAM-CVF spectrum (corrected from the dust continuum emission). \\ 
\end{table*}

\subsection{Comparison with observations}
\label{comparison_obs_mod}

We now compare these model predictions to the observations.
In Fig. \ref{comp_obs_mod}, we show the observed emission profiles along the cut 1 going through the
brightest filament. The CO emission profiles have been extracted from the PdBI maps. 
In Table \ref{Table}, for the observed CO emission lines and intensity ratios
we use the PdBI data when they are available and the IRAM-30m data otherwise. 
For the purpose of comparison of the observations with the models, we need to determine the position
of the PDR edge. The ionisation front is the {\it natural} edge of a PDR but in practice marking it is hard. 
We choose $d$=0 where the H$\alpha$ intensity is decreased by 50\%.

From comparison with observations,
we find that  :

- the H$_2$ emission is over-estimated and peaks too close to the edge
 for both values of $n_{H}$. This discrepancy is much more important for $n_H=2~10^5$ cm$^{-3}$ which 
produces a higher H$_2$ line intensity by three orders of magnitude and a much narrower
H$_2$ filament width. Also, this model produces the H$_2$/PAH intensity ratio 
much higher by a factor of the order of 20. 
For $n_H$=2 $10^4$ cm$^{-3}$,  the H$_2$ line intensity is over-estimated by a much lower factor (about 4)
 and the H$_2$/PAH intensity ratio
is roughly comparable to the observations.

- the CO emission peaks much too far inside the cloud for $n_H$=2 $10^4$ cm$^{-3}$.
Moreover, for this gas density value the C$^{18}$O lines emission and the C$^{18}$O/$^{12}$CO (2-1) line intensity ratio
 are underestimated by a factor of 5 and 2, respectively. 
On the contrary, for $n_{H}=2~10^5$ cm$^{-3}$ the peak position of the C$^{18}$O emission as well as the CO intensities and intensity ratios are  roughly well reproduced.

\par\bigskip
In summary, we find that models with a constant gas density equal to $n_{H}=2~10^4$ -- 2 $10^5$ cm$^{-3}$ cannot reproduce the global spatial stratification observed as well as the intensities of both the H$_2$ and CO lines.
One explanation of the discrepancies is that the Horsehead
nebula PDR presents at the edge a gas density gradient with $n_H\sim 10^4$ and $10^5$ cm$^{-3}$ in the H$_2$ emitting and inner cold molecular layers, respectively.
In the next section, we investigate what models with a gas density gradient profile predict.

\subsection{Gas density gradient}
\label{gradient}

We consider models with a gas density gradient profile defined as :
\begin{equation}
n_H(d) = \left\{ \begin{array}{r@{\quad:\quad}l}
n^0_{H} \times (d/d_0)^{\beta} & d \le d_0 \\ n^0_{H} \hfill  & d > d_0
\end{array} \right.
\label{Eq:densite}
\end{equation}
with $d$ the distance from the edge of the PDR and $n^0_{H}$ the constant gas density in the $d > d_0$ region. We take $n^0_{H}$=$2~10^5$ cm$^{-3}$ since high value of $n_H$ is required to account for the 
observed C$^{18}$O emission profile, as discussed above.
For the scale length ($d_0$) and the slope ($\beta$) of the gradient, we adopt $d_0=$0.02 pc (or 10'') corresponding to the peak of the 1.2 mm dust
continuum emission and $\beta$=1-4.

\begin{figure}[htbp]
\leavevmode
\begin{minipage}[c]{7cm}
\centerline{ \psfig{file=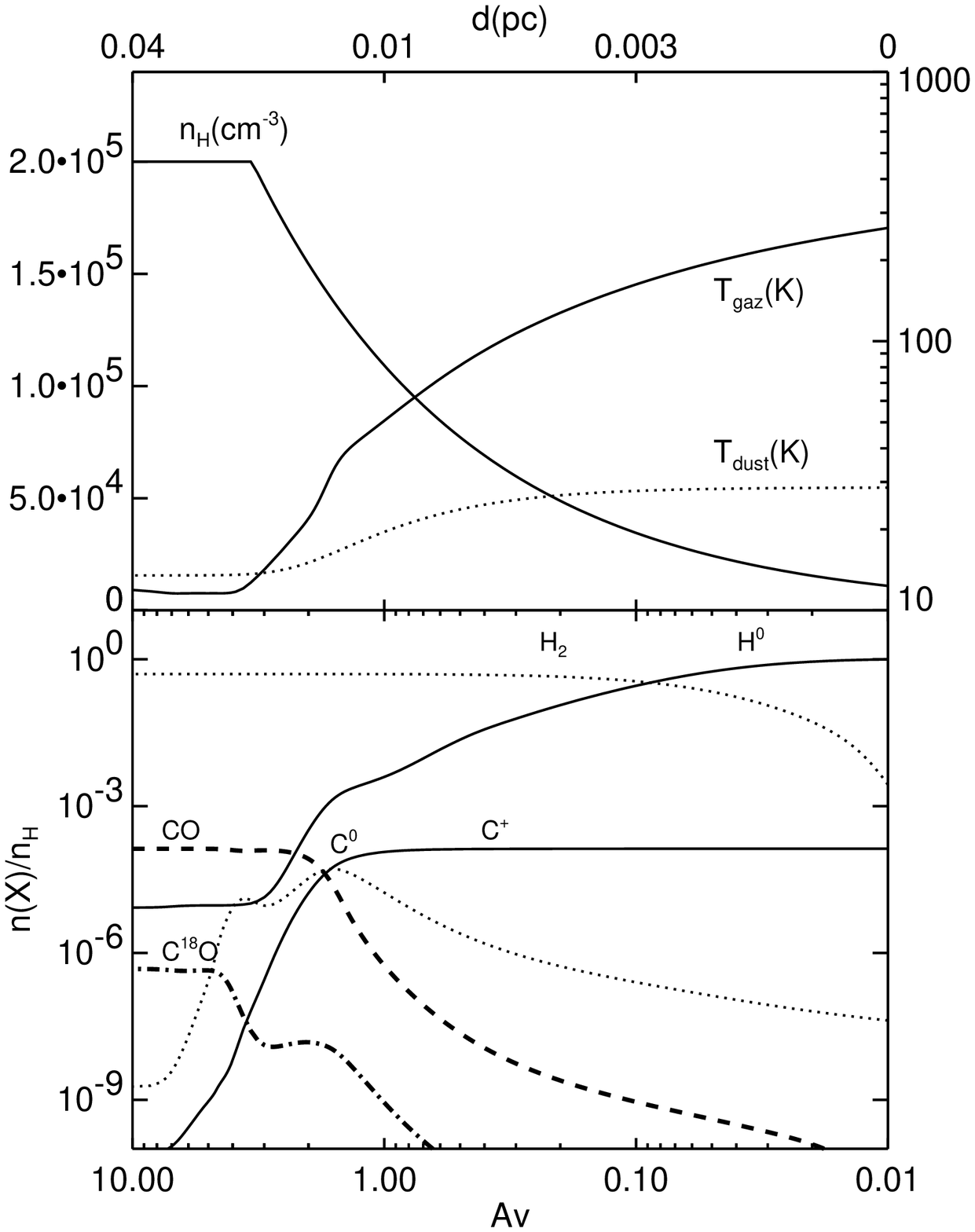,width=7cm,angle=0} }
\end{minipage}
\begin{minipage}[c]{7cm}
\centerline{ \psfig{file=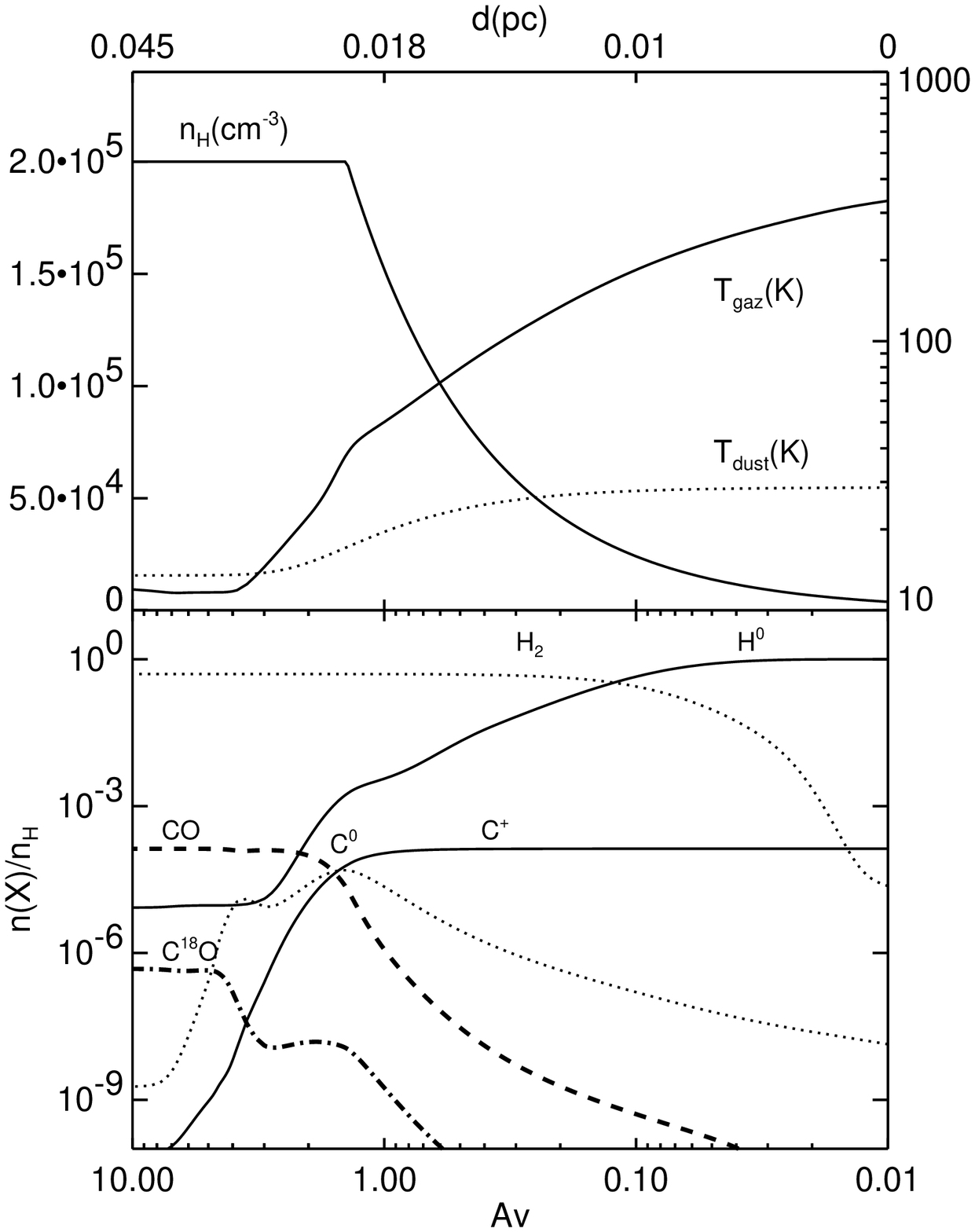,width=7cm,angle=0} }
\end{minipage}
\caption{\em
Same as in Fig. \ref{param1} for a model with a gas density gradient with $\beta=1$ (upper panel) and $\beta=4$ (lower panel).}
\label{param3}
\end{figure}

\begin{figure}[htbp]
\leavevmode
\centerline{ \psfig{file=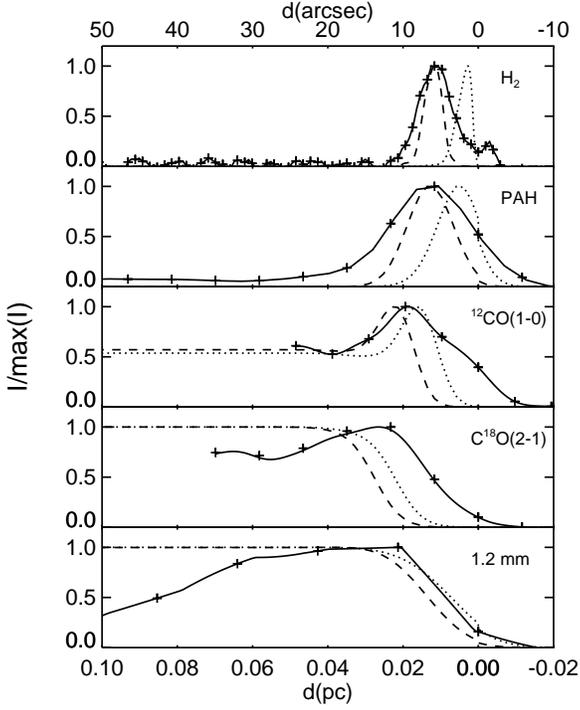,width=10cm,angle=0} }
\caption{\em  
Same as Fig. \ref{comp_obs_mod} for models with a gas density
gradient with $\beta=1$ (dotted lines) and $\beta$=4 (dashed lines). For more details
see text in Sect. \ref{gradient}. 
}
\label{comp_obs_mod2}
\end{figure}

In Fig. \ref{param3}, we show the model predictions of the thermal and chemical stratification as a function of the depth into the PDR and in Fig. \ref{comp_obs_mod2} the corresponding emission profiles. In Table \ref{Table}, we give the corresponding 
intensities and intensity ratios. 
It clearly appears  that the models with a gas density gradient reproduce
 much better the observations. 
In fact, for these two models, we have $n_{H}\sim 10^4$ and $10^5$ cm$^{-3}$ in the outer and inner region, respectively, and the emission profiles of
both the H$_2$ and CO lines are found to be roughly similar to the observed ones. 
Moreover, these models predict intensities of both the H$_2$ and CO lines
emission, as well as the intensity ratios, comparable to the observations.

For both models considered here, one can see that the scale length of the gas density
gradient ($d_0$) corresponds to that of the gas temperature change.
$T_{gas}$ decreases in fact from $T^{max}\sim$200-300 K to $T^{min}\sim$10-20 K on a scale length of the
order of $\sim$0.02 pc (see Fig. \ref{param3}).
At the edge where $n_H$ is low the gas is $\sim$10 times warmer than in the inner region 
where $n_H$ is 10 times higher.
 Also, we find that for these models the thermal pressure is roughly constant throughout the PDRs 
and equal to $\sim 4~10^6$ K cm$^{-3}$. 
This suggests that the thermal pressure dominates the gas pressure.

\begin{figure}[htbp]
\leavevmode
\centerline{ \psfig{file=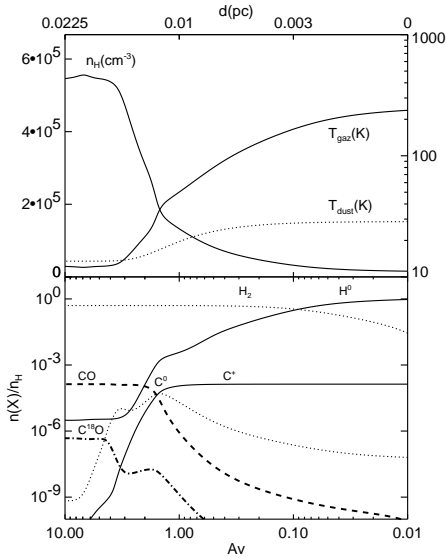,width=7cm,angle=0} }
\caption{\em
Same as in Fig. \ref{param1} for a model with a uniform pressure equal to $4~10^6$ K cm$^{-3}$.}
\label{param5}
\end{figure}

In order to test this interpretation, we run a model with a uniform pressure equal to $4~10^6$ K cm$^{-3}$
as found above.
In Fig. \ref{param5}, we show the model predictions of the thermal and chemical stratification as a function of depth into the PDR.
The gas density goes from $\sim 10^4$ cm$^{-3}$ in the H$_2$ emitting region
to  $5~10^5$ cm$^{-3}$ in the inner cold molecular layers
and the thermal and chemical structures are very similar to
  that of the gas density gradient models (see Fig. \ref{param3}).
For the isobaric model, the scale length of the gas density and temperature gradients
is also of the order of $d_0\sim 0.02$ pc.
Consequently, we find that the corresponding emission profiles
 and intensities are in good
 agreement with the observations as for the gas density gradient models.

\par\bigskip
Nevertheless, we note that for both gas density gradient and isobaric models considered here there are several discrepancies
with the observations:

- the H$_2$ emission profile is narrower than
the observations and the H$_2$ intensity is overestimated by a factor of 2-4.  Moreover, the  H$_2$ and PAH emissions have a tendency to peak too close the edge. 

- the C$^{18}$O line emission is shifted too much in the inner
region by $\lesssim 0.01$ pc (or $\lesssim$ 5'').

However, projection effects along the line of sight may be important
 and could affect the width and the position of the emission profiles, as well as, the absolute intensities.
In the following, we investigate the impact of the projection effects resulting from PDR inclination and 
sub-structures.

\nopagebreak

\nopagebreak

\subsection{Projection effects}
\label{projection}

\begin{figure}[htbp]
\leavevmode
\centerline{ \psfig{file=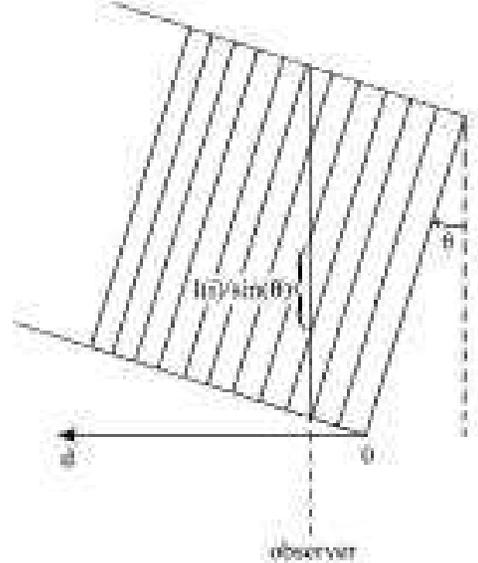,width=7cm,angle=0} }
\caption{\em Schematic diagram of a PDR inclined by an angle $\theta$ relative to the line of sight.
Index $j$ lables the $A_V$-layers of the model.}
\label{schema}
\end{figure}

First, we discuss the effects of PDR inclination on the emission profiles.
For a PDR tilted by an angle $\theta$ relative to the line of sight,
the emission integrated along a given line of sight $i$ 
 can be written as:
\begin{equation}
I(i)=\sum_{j=i}^{j_{max}} \frac{J(j)}{4\pi} \times l(j)/sin(\theta)
\end{equation}
where $j$ labels the various layer crossed by the line of sight (see Fig. \ref{schema}),
$J(j)$ is the line emissivity extracted from the plan-parallel model at the $A_V(j)$-layers,
 $l(j)/sin(\theta)$ the length of layer $j$ along the line of sight and $j_{max}$ is defined as 
$\sum_{j=i}^{j_{max}} l(j)/sin(\theta) = l_{PDR}/cos(\theta)$.
For the CO line emission, $J(j)$ is multiplied by $\beta (j)$ the escape probability 
of the layer $j$ along the line of sight.

To examine the effect of the inclination, we calculate the maximum angle of inclination $\theta _{max}$ considering that the PDR edge presents a wall
of density. Using the observations
of H$_2$, $\theta _{max}$ can be given by $sin^{-1}(d_{H_2}/l_{PDR})$ with $d_{H_2} \sim 0.01$ pc the distance  of the H$_2$ emission peak from the PDR edge. 
Thus, taking $l_{PDR}=0.1$ pc, we find $\theta _{max}\sim$6$^{\circ}$.

\begin{figure}[htbp]
\leavevmode
%gradient beta=1: Int_obs_mod_incl_Lebx_d1_im2_b0_chi.ps 
%gradient beta=4: Int_obs_mod_incl_Lebx_d1_im2_b2_chi.ps
%isobare: Int_obs_mod_incl_Lebx_iso_pm4_chi.ps
\centerline{ \psfig{file=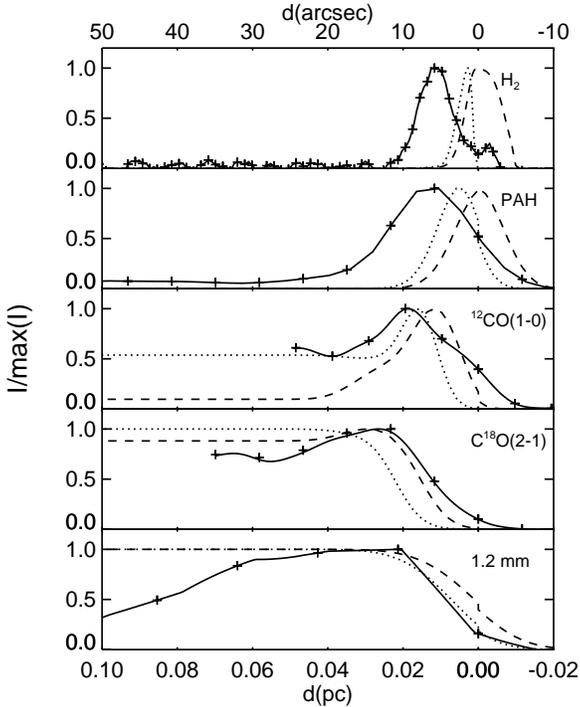,width=10cm,angle=0} }
\caption{\em  
Same as Fig. \ref{comp_obs_mod} for models with a gas density
gradient with $\beta=1$ (dotted lines) and taking into
account the inclination effects with an angle $\theta=6^{\circ}$ (dashed lines). }
\label{comp_obs_mod4}
\end{figure}

In Fig. \ref{comp_obs_mod4}, we show the results for the gradient density model
with $\beta$=1 inclined by $\theta =6$$^{\circ}$.
As expected, the gas and dust emission profiles are displaced further near the edge (by $\sim$0.005 pc or 2.5''). Moreover, the H$_2$ emission profile is larger (FWHM $\times$ 2) and the H$_2$ intensity is reduced by a factor of 2 (see Table \ref{Table}). The PAH emission, which is more extended, is on the contrary
practically not affected and the H$_2$/PAH emission becomes similar to the observed one. Finally, note that for the other models considered here
the inclination effects are found to be similar.

These results show that the inclination effects could explain the discrepancies 
between the observations and the gas density gradient and/or isobaric
models. 
Nevertheless,  taking into account these effects the H$_2$ and PAH emissions peak still too close to the edge.
This discrepancy could be in part due to the projection effects resulting
from the filamentary sub-structures observed at the edge of the Horsehead nebula (see Sect. \ref{h2_halpha_lw2}). 
In fact, for the comparison with the model, we have selected 
a cut across the interface where the two filaments appear closest but probably they are not precisly aligned
along the line of sight.
Furthermore, the position of the PDR edge is not really well determined
and it also has a slant.

\nopagebreak

\section{Discussion}
\label{discussion}

In summary, we conclude that in order to explain the observed spatial stratification of the different tracers, as well as their intensities and intensity ratios, a high gas density gradient at the edge of the Horsehead nebula  is required.  The scale length is about $d_0\sim$0.02 pc (or 10'') and the value of $n_H$ is about $10^4$ and $10^5$ cm$^{-3}$ in the H$_2$ emitting and inner cold molecular layers, respectively.
Since the scale length of the gradient corresponds to that of the gas temperature change ($T_{gas}$ decrases from $\sim$200-300 to 10-20 K on $\sim$0.02 pc) and
 the thermal pressure is roughly constant throughout the PDRs, 
we conclude that the gas density is given by the thermal pressure.
A model with a uniform pressure equal to $\sim$4 $10^6$ K cm$^{-3}$ fits in fact also well the observations.

This study is consistent with the earlier findings of such a gas density gradient on the western edge
of the $\rho$ Ophiuchi molecular cloud by \cite{habart2001,habart2003a}.
For this interface ($\chi \sim$250), we find from the analysis of the observed profiles of the H$_2$ 1-0 S(1), C$^0$ $^3$P$_1$-$^3$P$_0$
and $^{13}$CO(3-2) emission lines that the gas density
profiles is as in Eq. \ref{Eq:densite} with $n_H^0\sim$4 $10^4$ cm$^{-3}$, $\beta \sim$2.5 and $d_0 \sim$0.05 pc corresponding to the  scale length of the
gas temperature change.

\par\bigskip
Here it must be emphasized that our results are based on several important assumptions.\\
\noindent
{\it i)} We assume $\chi$=60 while $\chi$ could be lower.
However, a large angle between the sky plane and the line connecting the Horsehead to the star
 seems unlikely (see Sect. \ref{horsehead} and \ref{h2_halpha_lw2}) and, even
 diminishing the value of $\chi$ by a factor of 2 will not cause significant differences. \\
\noindent
{\it ii)} We assume that the H$_2$ formation rate, $R_f$, is equal to the standard
rate derived in the diffuse ISM while it could be higher. In \cite{habart2004}, we show that
for moderately excited PDRs (i.e., $\chi \le$1000), such as the Horsehead nebula, the
 H$_2$ formation rate is a factor of $\sim$5 times larger than the standard rate, $R^0_f$.
However, we have run models with $R_f=5 \times R^0_f$ and find that
for the conditions prevailing in the Horsehead nebula the chemical and thermal structure is practically not affected.
Nevertheless, note that the H$_2$ 1-0 S(1) line intensity will be increased. 
However, as our results are mainly based on the analysis of the spatial stratification, we conclude that the assumed value for $R_f$ 
does not greatly affect our conclusions.\\
\noindent
{\it iii)} We adopt $k_{\nu}$ constant and equal to 0.0035 $\rm{cm}^2~\rm{g}^{-1}$ 
which has an uncertainty of roughly a factor 2 \cite[]{bianchi2003}.
However, a change in the value of $k_{\nu}$ by a factor of 2 will only modify $l_{PDR}$ and the line intensities by at most a factor 2.\\
\noindent
{\it iv)}
There are also two fundamental uncertainties in our present study.
One is that we assume a homogeneous gas density while 
PDRs could be clumpy. 
\cite{abergel2003} conclude that the material behind the illuminated
edge may be non homogenous with clump sizes of $\sim$10$^{-3}$ pc or smaller.
However, \cite{gorti2002} 
have recently investigated the effects of FUV radiation on clumps and show that in PDRs, the ambient FUV penetrates through the surface of a dense clump and heats this surface to high temperatures, causing mass loss and thereby inducing photoevaporation of the clump. 
Smaller clumps are shorter lived compared to larger clumps 
and the photoevaporation timescale would be as short as $\sim 5~10^3$ yr for a clump size of 10$^{-3}$ pc and an average mass of $\sim 10^{-3}$
M$_{\odot}$ (or a density of $\sim 10^7$ cm$^{-3}$). \\
\noindent
{\it v)}
Another assumption is that of a static, equilibrium PDR. 
In reality, the propagation of 
the ionization and photodissociation fronts will bring fresh H$_2$ into the zone emitting line radiation.
The effects of a moving ionization front become very important for the PDR structure
if the timescale for H$_2$ to flow across the PDR is smaller than the timescale for H$_2$ photodissociation,
leading to the merging of the dissociation front with the ionization front.
For a column density of H$_2$ from the PDR edge of $N(H_2)\sim 10^{21}$ cm$^{-2}$ (which is typically our case), the criteria for merged the dissociation front to the ionization front is \cite[]{hollenbach99}:
$v_{IF} \gtrsim 3~(\chi/n_H)$ km s$^{-1}$ with $v_{IF}$ the velocity of the ionization front.
Apparently, the condition is not met in the Horsehead nebula, since the
observed layered structure shows that the H$_2$ offset from the ionization front (see Fig. \ref{cut_sofi}). 
Qualitatively, the merger tends in fact to occur for PDRs heated by hotter stars with a high ratio of extreme-UV to FUV photons \cite[]{bertoldi96}.
Nevertheless, it is possible that the dissociation front is shifted closer to the ionization
front because of the advection of H$_2$.
However, \cite{stoerzer98} have modelled the PDR structure assuming an ionization
front moving into the PDR and found that non equilibrium effects are probably minor in objects similar to the Orion Bar,
where the ionization front propagation was expected to have significant impacts.
We conclude that presently, it is reasonable to use  stationary PDR models
in order to estimate the physical conditions.

\section{Conclusion}
\label{conclusion}

The aim of this study has been to determine the gas density structure of  the Horsehead nebula PDR
viewed edge-on using the observed emission profiles of different tracers.
We have obtained high angular resolution ($\sim$1'') imaging observations of the H$_2$ 1-0 S(1) line emission,
using SOFI at the NTT, in order to resolve the PDR structure in the outer zone.
Analysis of the H$_2$ fluorescent emission, very sensitive to both the FUV radiation field and the 
gas density, in conjunction with the aromatic dust and the H$\alpha$ line emission,
brings strong constraints on the illumination condition and the gas density in the outer PDR region.
Moreover, in order to trace the penetration of the FUV radiation field into the cloud and probe the gas 
density structure throughout the PDR, we have combined the infrared emission of H$_2$ and aromatic dust with millimeter observations of CO and dust 
continuum emission (obtained at the PdBI and the IRAM-30 m telescope).

From comparison with PDR model calculations, 
we show that a constant-pressure PDR model with $P\sim$4 $10^6$ K cm$^{-3}$ best fits the 
global spatial stratification observed, as well as, the absolute intensities and intensity ratios
of the different molecular lines and dust emissions.
Constant density models fail to reproduce all the available data at the same time.
A high gas density gradient with a scale length of about 10'' (or 0.02 pc) and $n_H$ rising from $\sim 10^4$ cm$^{-3}$ in the H$_2$ emitting region
to $\sim 10^5$ cm$^{-3}$ in the shielded cloud interior
is required to explain the observations.
The gradient scale length corresponds to that of the gas temperature change ($T_{gas}$ decrases from $\sim$200-300 to 10-20 K on $\sim$0.02 pc); the gas density gradient model corresponds to a constant-pressure model and we 
 conclude that the PDR is in pressure equilibrium (at $P\sim$4 $10^6$ K cm$^{-3}$).

To the best of our knowledge, this study represents the first observational evidence of the dominant
influence of the thermal pressure on the gas density structure of a PDR.  
Experimental evidences of the thermal pressure domination are extremely scarce up to now 
because of the small spatial scales involved.
The constraints derived here on the gas density profile are important for the study of physical and chemical
processes in PDRs and provide new insight into the evolution of interstellar clouds.

\acknowledgements{This paper was based on observations obtained at the European Southern Observatory
(ESO, Chile), Institut de Radio-Astronomie Millim\'etrique (IRAM) and with the Infrared Space Observatory (ISO). IRAM is supported by INSU/CNRS (France), MPG (Germany) and IGN (Spain). ISO is an ESA project with instruments funded by ESA Member States and with the participation of ISAS and NASA. 
The authors are grateful to M. Gerin and E. Roueff for fruitful discussions and their relevant comments and suggestions.}

%\bibliographystyle{/home/anteo/habart/gabi/natbib}
%\bibliography{/home/anteo/habart/biblio/all}

\bibliographystyle{natbib}

\end{document}